# Financial Services, Economic Growth and Well-Being
*A Four Pronged Study of the Uncertainty Principle of the Social Sciences, the Responsibilities of Fiscal Janitors, the Need for Smaller Organizations and Redirecting Growth that Generates Garbage*

**Ravi Kashyap**
**Gain Knowledge Group**
**City University of Hong Kong**
**October 2, 2013**

Edited Version: Kashyap, R. "Financial Services, Economic Growth and Well-Being." Indian Journal of Finance, Vol. 9, No. 1 (2015), pp. 9-22.

## Table of Contents

# I. Abstract

The primary topic of consideration here is the relationship between the Financial Sector and Economic Growth. This is, of course, a natural extension of studies that focus on various variables that influence economic growth and falls under the wider category of beneficial factors and policies aimed at increasing the welfare or well-being to society. It is worth considering this wider goal, at the outset, since the effective functioning of the financial sector is critically dependent on obtaining a deeper understanding of these factors. These factors will be broadly classified into what we refer to as the four prongs and will be elaborated upon in subsequent sections.

***The four prongs are like the four directions for an army general looking for victory and any attempt at financial service reform that does not consider all the four prongs will prove to be insufficient and will be incomplete at best.***

We need to consider all the four prongs because the first one tells us about the limitation of any relationships we uncover; the second tells us about what is the overriding need of the financial sector and whether we are deviating from the intended goals; the third tells us that keeping complexity in check is important for accomplishing our objectives; and the fourth one tells us where unintended growth or outcomes, that will provide no real benefit, can result, despite the care we take to adhere to the stipulations of the first three. Just like the four directions, we need to be aware that there is a degree of interconnectedness in the below four prongs.

1. The Uncertainty Principle of the Social Sciences

2. The Responsibilities of Fiscal Janitors

3. The Need for Smaller Organizations

4. Redirecting Growth that Generates Garbage

It is also important to gain a quick, yet more profound, comprehension of welfare and delineate its components into those that result from an increase in goods and services, and hence can be attributed to economic growth, and into those that are not related to economic growth but lead to a better quality of life. The reasoning here being that economic growth alone is an inadequate indicator of well-being.

Hand in hand with a better understanding of the characteristics of welfare, comes the need to consider the measures or metrics we currently have that gauge economic growth and supplement those with factors that capture well-being more holistically. This is important because, there would be little sense in pursuing policies aimed at increasing some widely used metric like Gross Domestic Product, GDP, if such policies do not lead to an increase in welfare and worse still, if they lead to an unintentional decrease in well-being; on a lighter note, it is worth pondering about which meaning of gross is applicable in the context of GDP. With this, we would look to either constructing new metrics all together or looking at ways to improve or supplement existing ones.

The ensuing discussion, with the use of several illustrative analogies, is meant to intuitively demonstrate the validity of the four prongs. A formal approach towards proving these will be outlined in the subsequent sections and augmented or amended in later versions of this study.

## II.    Discussion of the Four Prongs

Let us now turn to the different facets of the financial services sector and welfare that will have a significant bearing on our discussion. The financial services sector exists to facilitate the creation and transfer of goods and services. This happens through the medium of money and other financial instruments; which we will henceforth, collectively refer to as money equivalents. The adoption of these instruments helps to establish prices, which serve as a common denomination to aid the exchange of goods and services.

Let us look at some fundamentals that govern all financial instruments since it will shed light on the limitations of what we can hope to comprehend regarding the relationship between the financial sector and economic growth; and then delve into some of the nuances. It is also worthwhile to mention here that for most assertions made below, numerous counter examples and alternate hypothesis can be produced. These are strictly attempts at tracing the essentials rather than getting bogged down with a specific instance. However, any study requires forming a conceptual framework based on the more common observations, yet being highly attuned to any specifics that can stray from the usual. Also, for the sake of brevity, a number of finer points have been omitted and certain simplifying assumptions have been made. Given the scope and complications of the below discussion, drawbacks are hard to avoid and future iterations will seek to address these as they are discovered.

## 1. The Uncertainty Principle of the Social Sciences

The various financial instruments that exist today can be broadly viewed upon as vehicles for providing credit and a storage for wealth, for both individuals and institutions alike. The different instruments, both in terms of their nomenclature and their properties, then merely become manifestations of which and how many parties are involved in a transaction and the contractual circumstances or the legal clauses that govern the transaction.

Despite the several advances in the social sciences and in particular economic and financial theory, *we have yet to discover an objective measuring stick of value, a so called, True Value Theory*. While some would compare the search for such a theory, to the medieval alchemists' obsession with turning everything into gold, for our present purposes, the lack of such an objective measure means that the difference in value as assessed by different participants can effect a transfer of wealth. This forms the core principle that governs all commerce that is not for immediate consumption in general, and also applies specifically to all investment related traffic which forms a great portion of the financial services industry.

Although, some of this is true for consumption assets; because *the consumption ability of individuals and organizations is limited and their investment ability is not*, the lack of an objective measure of value affects investment assets in a greater way and hence investment assets and related transactions form a much greater proportion of the financial services industry. Consumption assets do not get bought and sold, to an inordinate extent, due to fluctuating prices, whereas investment assets will. The price effect on consumptions assets affects the quantity bought and consumed, whilst with investment assets, the cyclical linkage between vacillating prices and increasing number of transactions becomes more apparent.

***Another distinguishing feature of investment assets is the existence or the open visibility of bid and ask prices***. Any market maker for investment assets quotes two prices, one at which he is willing to buy and one at which he is willing to sell. Consumption assets either lack such an outright two sided quote; or it is hard to painlessly infer viewable buy and sell prices, since it involves some conversion from a more basic form of the product into the final commodity being presented to consumers. Examples for consumption assets are a mug of hot coffee, that requires a certain amount of processing from other rudimentary materials before it can be consumed; or a pack of raw almonds which is almost fit for eating. Coffee shops that sell coffee do not quote a price at which they buy ready drinkable coffee; the price at which a merchant will buy almonds is not readily transparent. Gold is an example of both, a consumption and an investment asset. A jewelry store will sell gold and objects made of gold; but it will also buy gold reflecting its combined consumption and investment trait. This leaves us with financial securities like stocks and bonds that are purely investment assets.

A number of disparate ingredients contribute to this price effect; like how soon the product expires and the frequent use of technology to facilitate a marketplace. EBay is an example of a business where certain consumption goods are being bought and sold. This can happen even if goods are only being sold, through the increased application of technology in the sales process. While not implying that the use of technology is bad, technology, or almost anything else, can be put to use that is bad. Thankfully, we are not at a stage where Starbucks will buy and sell coffee, since it can possibly lead to certain times of the day when it can be cheaper to have a cup of coffee and as people become wary of this, there can be changes to their buying habits, with the outcome that the time for getting a bargain can be constantly changing; making the joys of sipping coffee, a serious decision making affair. Even though this is an extreme example, we will overlook some of these diverse influences for now, since our attempt is to exemplify the principal differences between the varieties of financial transactions and the underlying types of assets that drive these deals.

The lack of an objective measure of value, (hereafter, value will be synonymously referred to as the price of an instrument), makes prices react at varying degrees and at varying speeds to the pull of different macro and micro factors. The greater the level of prevalence of a particular instrument (or even a particular facet of an instrument) the more easily it is affected by macro factors. This also means that policies are enforced by centralized institutions, (either directly by the government or by institutions acting under the directive of a single government or a coalition of governments), to regulate the impact of various constituents on such popular instruments. Examples for this would be interest rate dependent instruments, which are extremely sensitive to rates set by central banks since even governments issue such instruments; dividends paid by equity instruments which are clearly more sensitive to the explicit taxation laws that govern dividends than to the level of interest rates; and commodities like oil, which are absolutely critical for the smooth functioning of any modern society and hence governments intervene directly to build up supplies and attempt to control the price.

Lastly, it is important that we lay down some basics regarding the efficiency of markets and the equilibrium of prices. Surely, a lot of social science principles and methodologies are inspired from similar counterparts in the natural sciences. A central aspect of our lives is uncertainty and our struggle to overcome it. Over the years, it seems that we have found ways to understand the uncertainty in the natural world by postulating numerous physical laws.

These physical laws are deductive and are based on three statements - a specific set of initial conditions, a specific set of final conditions and universally valid generalizations. Combining a set of generalizations with known initial conditions yields predictions; combining them with known final conditions yields explanations; and matching known initial with known final conditions serves as a test of the generalizations involved. The majority of the predictions in the physical world hold under a fairly robust set of circumstances and cannot be influenced by the person making the observation and they remain unaffected if more people become aware of such a possibility.

In the social sciences, the situation is exactly the contrary. A set of initial conditions yielding a prediction based on some generalization, ceases to hold, as soon as many participants become aware of this situation and act to take advantage of this situation. This means that predictions in the social sciences are valid only for a limited span of time and we cannot be sure about the length of this time, since we need to factor in constantly the actions of everyone that can potentially influence a prediction, making it an extremely hard task.

***All attempts at prediction, including both the physical and the social sciences, are like driving cars with the front windows blackened out*** and using the rear view mirrors, that give an indication of what type of path has been encountered and using this information to forecast, what might be the most likely type of terrain that lies ahead for us to traverse. The path that has been travelled then becomes historical data that has been collected through observation and we make estimates on the future topography based on this. Best results generally occur, when we combine the data we get in the rear view mirror with the data we get from the side windows, which is the gauge of the landscape we are in now, to get a better comprehension of what lies ahead for us. The quality of the data we gather and what the past and the present hold then give an indication to what the future might be. So if the path we have treaded is rocky, then the chances of it being a bumpy ride ahead are higher. If it has been smooth, then it will be mostly smooth. Surely, the better our predictions, the faster we can move; but then again, it is easy to see that the faster we travel, the more risk we are exposed to, in terms of accidents happening, if the constitution of the unseen scenery in front of us shifts drastically and without much warning.

A paramount peculiarity of the social sciences is that passage on this avenue is part journey and part race. The roads are muddy, rocky and more prone to have potholes. This means being early or ahead on the road brings more winnings. We also have no easy way of knowing how many people are traveling on this path, either with us, ahead of us or even after us. As more people travel on the path, it starts falling apart, making it harder to travel on it, a situation which is accentuated considering we don't have any vision out front. On the other hand, let us say, physical science roads, being well paved and well-constructed using concrete, hold steady for much longer time durations, so what has been observed in the past can be used to make durable forecasts that hold for lengthier amounts of time in the future.

This inability to make consistent predictions in the social sciences and the lack of an objective measure of value or a True Price Theory means that is almost impossible for someone to know what a real state of equilibrium is. The efficient market hypothesis in spite of being a very intriguing proposition, can at best claim that markets have a tendency to move towards being efficient, though a state of equilibrium is never fully attained since no one has an idea what that state of equilibrium is and the actions of the participants serves only to displace any state of equilibrium, if it did exist. The analogy for this would be a pendulum with perpetual motion; it swings back and forth around its place of rest with decreasing amplitude and the place of rest keeps changing with time, starting a new cycle of movement with reinforced vigor.

We can then summarize the above with the ***Uncertainty Principle of the Social Sciences***, which can be stated as, ***"Any generalization in the social sciences cannot be both popular and continue to yield predictions, or in other words, the more popular a particular generalization, the less accurate will be the predictions it yields"***. This is because as soon as any generalization and its set of conditions becomes common knowledge, the entry of many participants shifts the equilibrium or the dynamics, such that the generalization no longer applies to the known set of conditions.

All our efforts as professionals in the field of social sciences, will then be to contemplate upon uncertainty and uncover quasi-generalizations; understand its limitations in terms of what can be the closest states of pseudo-equilibrium; how long can such a situation exist; what factors can tip the balance to another state of temporary equilibrium; how many other participants are aware of this; what is their behavior and how is that changing; etc., making our professions a very interesting, challenging and satisfying proposition.

With this in mind, we can turn specifically to how the above discussion applies to the relationship between the financial services and economic growth. As knowledge of a certain connection becomes available, it will attract the attention of many participants, who wish to benefit from it, leading to bubbles that eventually burst and the cycle of booms and busts continues. The extent of impact, in terms of countries or institutions affected, will depend on how prevalent the instrument or group of instruments, involved in the originally established connection, are. As we further investigate the services provided by the financial sector and its relationship to growth, we need to be cognizant of the temporal aspect of these changes; that the strength of the relationship will be changing continuously; and the upper limit in terms of the applicability of any patterns we unmask.

2. *The Responsibilities of Fiscal Janitors*

The importance of this prong lies in the proposition that once there is more clarity on what needs to be done; the how to do it part becomes relatively less obscure, making it more straightforward to figure out ways to accomplish it.

Water gives life and sustains it. It is required everywhere, for life, as we know it, to exist. In a similar vein, it is hard to imagine an economy without money or money-equivalents. This comparison is only partly valid, since life, as we know it, would cease to prevail without water, while we can essentially have a barter economy without money-equivalents. Barring this key limitation, the smooth functioning of a practical modern economy requires the flow of money-equivalents.

***Just as we have constructed devices and machines to control and divert the flow of water, to maximize the growth of life; we have the financial services sector that has to control and divert the flow of money-equivalents, to maximize the growth of an economy***. Taking the analogy a step further, our central reservoirs, irrigation canals, water tankers, pumping stations, pipes and water sprinklers are devised to keep water flowing around; similarly centralized and regional financial institutions, wire transfers, credit cards, checks, bank drafts, the internet and related technologies are meant to keep money-equivalents sloshing around.

We could construct different examples to elucidate instances where different amounts of water are required to fulfill various needs related to the growth of living creatures; likewise different amounts of money-equivalents are required to satiate various needs of an economy. Watering the garden is like running a small store or business; a large irrigation system is like running an enterprise, the size of a manufacturing plant; or a fire truck sending water to a fire is like giving funds to a business that is in distress and could go bankrupt.

Water is useful for many other purposes, like running industries that either directly or indirectly support life; while money, other than the odd instances we hear of someone using it for sleeping or bathing or as an item of decoration, is solely useful for economic activity. The uses of water for life like, cooking, drinking, cleaning, are similar to the uses of money for businesses like buying computers, paying salaries or rent. We can consider the different types of money-equivalents as different flavors of water pertaining to human life: bottled-water, soda or juice.

The people that come up with designs for devices that control and regulate the flow of water and keep it flowing are the engineers, hydrologists, water-plant operators, water-meter readers, plumbers, gardeners, farmers, fire-fighters, etc., collectively referred to as the water handlers. Similarly, the people involved in designing mechanisms for the flow of money-equivalents and maintaining the flow are the financial engineers, traders, bankers, accountants, lawyers and other financial-services staff, collectively referred to as the fiscal janitors.

Let us *imagine a situation where the civil engineers and other water handlers, have an insatiable appetite for water and instead of building a system that keeps the water flowing, wish to retain as much of the water for themselves, as possible*, without raising a lot of eyebrows. The result would be that the growth and sustenance of life elsewhere is thwarted. The systems they design to carry water would be non-optimal and instead of maximizing flow, it would have mechanisms built in, that would divert the water towards storage for themselves or the mechanisms would send water to places from which the chances of the water coming back to them are higher. In essence, the water would not reach the places where it is most needed, but it would reach places from where the likelihood of it flowing back or getting diverted for alternate storage is higher.

Imagine the consequences on the garden if the gardener, in return for services rendered, starts taking large swigs off the watering hose, equal in amount if not more, than the water tended to the plants; or an irrigation canal built to collect back or divert a high percentage of the water that is supposed to flow through it into the fields; what of a fire truck that will spray large amounts of water into a hidden storage tank instead of sending it entirely towards a raging fire. A micro analogy here can be regarding blood flow within organisms; imagine the aftereffects on the life of the organism, if the heart had a quenchless thirst for blood.

The natural system, of rainfall or snowfall; accumulation of water in lakes; snow on the mountains or ice in the ice caps and glaciers; the snow melt or lake fed flow of rivers into the sea; the process of evaporation; condensation and the forming of clouds; and the dispersion of clouds by the wind; is a watering network, we should someday hope to emulate and has no strict parallels for now in our economy. The largest store of water in the oceans, with its saltiness, renders it useless for most purposes, acting as a natural deterrent for anyone that wishes to hoard it. The question of whether we have as much wealth as there is water in the oceans is a can of worms better left unopened for now. A quote by Mahatma Gandhi can be modified to infer that, there is enough to satisfy every man's need, but not enough to satisfy one man's greed. Suffice it to say, without getting too philosophical, the Earth has enough resources to nourish all life on it. An affiliated concept is the expiration of money-equivalents. While inflation works like aging and weakens the value of money-equivalents, pragmatic methods of systematically perishing stockpiled reserves need to be explored.

This is a key problem of the financial-services industry. The people that are supposed to keep money-equivalents flowing around want to withhold as much of it for themselves as possible. Instead of finding ways to keep money circulating around, they are devising ways to attract and retain it. They are building money magnets as opposed to building ways to pump money and keep it streaming around. Once the flow starts getting diverted away from productive endeavors, a vicious cycle sets in, creating more opportunities for people to enter the business of diverting the flow, to make a living, rather than being the recipients of the flow and using it for other activities. The brightest people get drawn, perhaps involuntarily, to designing flawed systems that retain most of the flow, instead of sending it where it is needed most.

There are four ways to circumvent this issue; one, by having people in the financial-service industry that do not want to earn exorbitant compensations. This stricture is rather hard to assuage, since the financial industry, currently known for high salaries, tends to attract people that wish to earn huge sums of money. The claim that most people have an insatiable appetite for money and are eager to earn large sums, making this a harder option to implement, wouldn't be entirely untrue. Two, financial services could operate against the profit-maximizing ideal of businesses, with the danger that we might have large inefficient bureaucracies. Three, centralized institutions can keep printing new money, or introducing new money-equivalents, whenever the circulation dries up. This brings with it a whole host of other problems and again leads to more proliferation in the hands of those with the more powerful money magnets. Four, restrict the amount of flow that a certain financial institution has access to; since the lesser the flow, lesser will be the amount available to divert or retain. This premise naturally leads us to consider the need for smaller financial organizations.

## 3. *The Need for Smaller Organizations*

The water circulation analogy highlights the need for organizations that are meant to be pumping money around to be small, since they could then keep less of it for themselves. Smaller size also means that it would be easier to check what the institutions and the people involved are doing. We need to look at the argument that if organizations are small, there would be many such institutions, making it a harder task to monitor them.

With the reduced scope of smaller financial institutions, there would be a stronger relationship between the service provider and the served. The smaller size leads to more number of superior quality interactions between the same parties, leading to a repeated game setting, which produces more co-operative behavior. ***The strengthened relationship effectively acts as an enforcement agent towards both the parties.*** The smaller institution cannot extract large rents for itself, since otherwise it would cease to be competitive against the myriad number of smaller institutions that are vying for business and such actions would deter people from doing business with it. The people that are benefiting from the services of an institution would be under close scrutiny from the institution itself, which bears the burden of ensuring that its services are put to the best use possible, since that is integral for its own survival. Agents are driven against myopic self-motivated behavior, since maximal benefits accrue by acting with a longer-term vision.

***Smaller size reduces complexity in many ways and makes it harder to hide things under the rug.*** This makes it tougher for corruption or other illegal episodes to happen. Systemic failures, wherein most organizations in a sector, are severely affected in a negative way, are less likely, since we have many small organizations and the degree of interconnectedness will be lower. The actual provision of service through different forms of money-equivalents and questions regarding which one is better suited become a secondary concern, since they are simply different contractual terms, involving different parties. ***Any set up where the players involved have a fundamental incentive to be on best behavior, functions better than other alternate possibilities.***

The analogy about the insatiable appetite for money applies to all organizations, not just financial ones. Instead of being immersed in a pursuit that maximizes a particular kind of output, the organization and its people will engage in practices that are aimed at collecting the most amounts of money-equivalents. The current implicit consensus on this front is that maximizing the collection of money-equivalents will produce the peak amount of output. The associated growth in output might or might not end up as showing the maximum amount of GDP; even if we can make a case that it will result in the highest amount of GDP growth, we can show that it will not result in the most amount of growth that is favorable to society and does not bring welfare to the most number of people.

As organizations grow bigger, a greater proportion of the individuals that are part of it, become involved in just making sure things are running smoothly. This takes people away from becoming involved with the actual generation of ideas or producing a tangible output or adding to real growth. This is especially true in the financial-services sector, where the amount of innovation is empirically known to be lesser than other sectors. Innovation and Intelligence, despite their importance, are less important than Integrity in financial-services; there is no need for geniuses to make loans or conduct other financial transactions, we simply need more transparency, which will result in more fairness and the right thing being done. Honesty is not entirely innate; it can be instilled and it follows from the recognition that human conduct is usually a response to the incentives and the situations. While, formal attempts at tracing the impact of integrity on the functioning of financial institutions are worthwhile; a simpler argument, that smaller organizations with less complexity, create a better alignment of incentives, and give rise to an environment where it is harder to hide immoral incidents, and foster more righteous behavior, can be shown to hold water.

We could raise the point that the compensation of executives in large firms can be monitored and going this route would be easier than having to monitor thousands of smaller firms. The rebuttal for this would be that when someone has access to large amounts of money, the chances of misappropriation are higher than when there is no access to large amounts. The recent financial crisis had instances where large bonuses were paid out even by firms that were receiving bail out funds from the government, under the excuse of retaining talent, among others. While it is not entirely inappropriate to impose limits on executive compensations, it is highly likely that clever modes of excessive compensation will be devised, when there is access to siphon large amounts of money. The story of Sergey Bubka, the pole-vaulter who broke the world record thirty-five times, illustrates the limits of raising the regulatory bar. Each time the pole was raised, he would jump higher. This is also about the system of governance. Simply put, should the state should interfere with the specifics of how a firm is run or should the state restrict the main activities of a firm. The later sections consider this in further detail, but without digressing much farther, we can surmise that giving the state power over every day affairs can be disastrous.

*Does size matter?* Does large size lead to stability? Turning to nature again for inspiration, we don't see excessively large organisms, despite some creatures that never stop growing. Similarly, organizations have a tendency to grow. We are a growth obsessed society. A mindset that tolerates the omnipresent stressor of competition and celebrates the birth and death of organizations helps prevent abnormal growth. The pseudo-stability of big organizations can cause disasters when they fail, since most systems can cope better with many small continuous demises than a few large sudden deaths. ***Size does matter.***

Bigger organizations could hide inefficient parts and subsidize their existence. The argument about economies of scale is not as applicable today, because, we use automation and machines extensively for agriculture and producing goods. Organizations are knowledge based as opposed to traditional manufacturing, for which such a production term needs to be applied. If organizations are to be small, it is helpful to have a climate that facilitates entrepreneurial activity, allowing the easy birth and growth of new businesses and large organizations do not stifle smaller ones.

Tracing the historical development of the trend towards bigger organizations, supported by bigger communities, leads us to the key stimulus that was the surplus generated by the superior modes of agricultural production. This made possible the establishment of a non-producing class, whose members were crucial for the rapid development of writing, science, cities, technology-based military prowess and formation of states. Dense population centers that could be supported near these lush agricultural centers had a greater exchange of ideas, bringing new innovations into force and allowing the extraction of rents from a larger number of individuals who came to depend on these new products that were fed by the invention spree.

The blessings of large population centers, on the sciences and the arts, have been tremendous. Development of regions like Silicon Valley in California or Broadway in New York is due to the rapid exchange of ideas. While the benefits of dense populations accrue up to a certain point, the negatives of overcrowding, shortage of resources and diminishing returns set in after a certain stage, giving rise to increasing disparities between the residents in these packed colonies. The widespread use of technology to connect people facilitates interaction among relatively far flung dwellings, removing the need for the congregation of individuals to accelerate the pace of evolution of human civilization.

This prong is the most important one, since if we get this right, the reduced size and complexity helps realize the limitations; aids in the detection of variations from the expectations; ensures that the responsibilities and incentives of the parties involved are aligned with the original targets and continue to stay aligned; and makes it easier to ascertain the unintended consequences of any efforts, which are hard to completely eliminate, as we will see next.

## 4. *Redirecting Growth that Generates Garbage*

The water propagation analogy alerts us to the aspect that flow could get diverted not where it is required most, but where certain parties might retain most of it. This could lead to growth that does not produce the most amount of welfare, which means, if all the growth is not leading to welfare there must be some growth generating garbage. A part of the growth, as we currently measure it, results from the back and forth flows of either water or money-equivalents, away from the intended destination, wasteful to the final goal of maximizing welfare. This can be compared to the trading of assets frequently, which does not result in real growth, but appears to give the illusion of progress. High frequency trading (HFT) then becomes a higher speed of water flowing back and forth, not helping growth happen in the places where it supposed to happen.

The point that requires further consideration is whether HFT is leading to benefits by either directly providing additional liquidity or indirectly via the spawning of numerous technological innovations in computer networking hardware, software or other items used to facilitate HFT, which can then be beneficial to other sectors. An example that concerns high speeds is from the car racing industry, which comes up with innovations that produce faster and safer cars. Many of these innovations slip into the mainstream automobile industry over time. Completely restricting any endeavor is not ideal since it hard to know where the next life changing idea might spring up; but regulating the dangerous or unfavorable ones is prudent. Learning further from this example, we do not see racecars cruising down our town streets (though, not something to rule out entirely in the not too distant future); they hustle around in an exclusive arena, indicating that perhaps we need a similar mandate for the HFT industry.

Another example regarding growth that does not add significantly to the welfare bottom-line is set in a hypothetical country where everyone is buying a certain kind of new cup, which, in addition to its benefits as a holder of fluids, looks nice and is bought just as a curio. This cup has become such a huge success, that people are buying it, in multiple quantities, just to keep it in their living room bookshelves along with their other collectibles. A lot of effort will go into making such cups and selling them. Dozens of copycat manufacturers enter this business and soon we have a thriving economy running on cup making and selling. This increased purchasing of the cup will raise the GDP growth in two ways, one directly through the higher number of cups being sold; the other through the increased price of the cups, as more demand pushes the price up. A safe assumption would be that other ancillary industries like graphic-design, material research etc. would spring up to support this cup making.

This continues for a while, the GDP and economic growth increase by leaps and bounds, eventually leading to a bubble, when the prices of cups reach a point, prompting the question, "Is this cup too expensive?" This is a rather involved topic since the state of pseudo-equilibrium, with increasing prices and selling of cups, can continue for long periods of time without the realization of being in a bubble. One indicator of a bubble is, if price increases in any sector have outpaced price increases in other sectors and the median wage growth of the corresponding population.

Some participants in the cup making industry, the manufacturers and sellers of the cups; the people that finance the growth of this industry by lending credit for new entrants and for subsequent expansions, accumulate a large share of money-equivalents. When the bubble bursts, the same economy will be selling fewer cups at lesser prices, and probably even show a recession. Given the simplified nature of the example, let us ignore that the money stashed by certain participants could get used in other areas, fuelling growth elsewhere. This again could be the growth that leads to welfare or the other kind of growth that generates garbage, both literally and figuratively.

Any claim that a collapse of this cup economy is having a severe negative effect on the lives of its citizens is dubious. We can consider three scenarios: one with a completely closed economy, with no exports or imports; one that is partly closed; and one that is entirely open. Under all three cases, ***the presence or absence of a few extra cups will not create a delightful paradise or lead to abject misery***. What would differ is the extent of deficit in the imports of food and other must-haves, fueled by the purchasing power of cup exports. Instead of a cup, our example works equally well with any number of items like cell-phone covers, drink-umbrellas, watch-straps or even cell-phones, or watches since ***the question of what is absolutely imperative to lead a good life is a constantly changing one, as luxuries end up becoming necessities***. If we take a moment to evaluate what really brings welfare, we can make the affirmation that welfare results mostly from the increased production of essential consumption goods and from the reduction of uncertainty in their procurement. Hence, any growth that is happening in certain sectors of the economy is more vital to the well-being of the masses than certain others.

To determine what is intrinsic to well-being, requires acknowledging its subjectivity. ***Poverty is a state of mind***. Layard (2010) talks about surveys in the United States that showed no increase in happiness over the past 60 years, reflecting the fact that higher national income has not brought the better quality of life that many expected. The science of subjective well-being is young, but it is developed enough to know that we need to collect data and make it a prime objective to quantitatively study the determinants of well-being, so that it can be used in policy analysis.

We need to scrutinize the metrics we have to measure welfare. Surely, GDP is insufficient and misleading. We need metrics that measure the distribution of consumption goods and supplement those with measures that gauge how quality of life improves. The level of health and education would be important. In addition, we need sound measurement of employment security; the amount of leisure spells people have; environmental factors relating to air, water, noise pollution; personal safety against crimes and conflicts; social factors like availability of support in case of need; freedom to express oneself; political participation and lastly, sustainable ways of production, which boils down to making sure that what we produce today can continue to be produced with minimal impact to the environment and being able to maintain the current level of well-being for future generations.

If governments or any organizations start intervening to set strict limits on production, on the amount and manner in which people consume and live, the results could be catastrophic and lead to too much state control or socialism or even communism. (Hayek 2009; Marx & Engels 1848; are seminal works). The other extreme is completely free markets, or capitalism, which as we are realizing will lead to huge inequalities in society. While it is hard to draw a strict line between these two modes of governance, we need elements from both models of governance and economic policy. Our earlier discussion on the advantages of smaller size firms applies here, wherein, with the prevalence of smaller organizations, incentives would be aligned such that it becomes easier to spot wrongful conduct, and identify deviations from intended outcomes, leading to better governance with lesser monitoring.

## III.    Literature Review

### 1. *Measure of Well Being*

There is a compelling case for constructing better metrics. The Stiglitz, Sen and Fitoussi Report (2009) highlighted the deficiencies in existing metrics, encapsulated an agenda for improvements, and discussed key areas on which further research is needed. We need to use some of the suggestions from this report to construct our measure of well-being.

### 2. *The Uncertainty Principle of the Social Sciences*

Paiche and Sterman (1992) inquire into decision making in complex environments and conduct an experiment where subjects must manage a new product from launch through maturity, and make pricing and capacity decisions. Building upon previous studies, they demonstrate that decision making in complex dynamic environments tends to be flawed in specific ways by not accounting sufficiently for feedback loops, time delays and nonlinearities. Even with a decent amount of experience, there is no evidence that environments with high feedback complexity can produce improved decision making ability.

The literature is rife with attempts of prediction, ranging in scope from small neighborhoods of a few dozen people all the way up to nations that are responsible for the welfare of millions. We need to concern ourselves with models that relate financial services, economic growth and welfare, though we can come up with interesting instances of pseudo-equilibrium states in market efficiency studies, relating to the price movements of some commonly traded financial instruments, since it would be a slight deviation from the central endeavor. We need to try to do this across different population sizes and how this varies across time. We will look at specific findings from various scholars in the next section. We need to reflect upon the temporal and population scope variations of the associated predictions in the later iteration or in attempts to add to this study.

### 3. *The Responsibilities of Fiscal Janitors*

We need to look at financial services activity that is leading to the formation of bubbles and also assess the development in complexity of the services offered by the sector. Philippon and Reshef (2008) focus on the human capital dimension of financial services. They look at skill composition and relative wages from 1909 to 2005 and find that deregulation has contributed to a significant increase in the wages and the need for skilled labor. They also show that corporate finance needs from the non-financial sector help explain the demand for skills in the financial industry. They contend that tighter regulation is likely to lead to an outflow of human capital out of the financial industry. Whether this is desirable or not depends on one's view regarding economic externalities. Murphy, Shleifer and Vishny (1991) and Philippon (2007) argue that the flow of talented individuals into law and financial services might not be categorically desirable, because social returns might be higher in other occupations, even though private returns are not.

Levine (2005, 1997), Arestis and Demetriades (1997), Levine and Renelt (1992), King and Levine (1993), Levine and Zervos (1998) discuss at length the impact of financial services on growth. They consider different facets like the impact due to stock markets, entrepreneurship, financial liberalization, international trade, fiscal policy, monetary & fiscal policy, political indicators and foreign direct investment. Existing research suggests that the relationship is not just of financial services following economic growth or vice versa and that there is much work required to study the co-evolution of growth and finance. There is also much work required on the relationship between financial structure and economic growth. What also remains to be explained more comprehensively is if, whether and to what extent a market with lower financial transaction costs promotes innovation and stimulates the invention of new and better production technologies.

Borensztein, Gregorio and Lee (1997) discuss the relative benefits of Foreign Direct Investment to growth through the transfer of more advanced technology. They bring to focus the importance of the development of human capital in the recipient country for the welfare maximizing absorption of the foreign investment.

The Sweeney and Sweeney (1977) anecdote about the Capitol Hill baby-sitting crisis exposits the mechanics of inflation, setting interest rates and monetary policies required to police the optimum amount of money. The creation of a monetary crisis in a small simple environment of good hearted people expounds that even with near ideal conditions, things can become messy; then in a large labyrinthine atmosphere, disaster could be brewing without getting noticed and can strike without much premonition. This emphasizes the need to keep complexity at bay and establishing an ambience where repeated games can be played with public transparency, so that guileful practices can be curtailed.

## 4. The Need for Smaller Organizations

Beck (2008) mentions that bank size is positively correlated with complexity so that large banks are harder to monitor than small banks. De-Nicolo (2000) argues that bank consolidation is likely to result in an average increase in banks insolvency risk. Beck & Demirguc-Kunt (2006) find that small firms face larger growth constraints and have less access to formal sources of external finance, potentially explaining the lack their contribution to growth.

Acs and Varga (2004) highlight two important proxy measures of the existence of entrepreneurial opportunity, the tendency of people to engage in self- employment and the tendency of people to start new firms. Using data from the Global Entrepreneurship Monitor (GEM) project they examine the relationship between entrepreneurship, knowledge spillovers and economic growth. There are manifold ways to measure entrepreneurial activity. One overbearing dissimilitude is between opportunity-based entrepreneurial activity and necessity-based entrepreneurial activity. Opportunity entrepreneurship represents the voluntary nature of participation and necessity reflecting the individual's perception that such actions presented the best option available for employment, but not necessarily the preferred option. Opportunity entrepreneurship differs from necessity by sector of industry and with respect to growth aspirations. Opportunity entrepreneurs expect their ventures to produce more high growth firms and provide more new jobs. Any measure of entrepreneurial activity needs to factor in this distinction. Bockstette, Chanda and Putterman (2002) construct an index that captures the length of state experience. This index is higher for countries that have a longer experience with state-level institutions and such countries have higher political stability, institutional quality and economic growth. We need to supplement the level of entrepreneurial activity with a measure that captures the depth and history of entrepreneurial culture present in a country.

Acs and Varga (2002) hypothesize that any spatialized theory of technology led regional economic growth needs to reflect three fundamental issues. First, it should provide an explanation of why knowledge related economic activities start concentrating in certain regions leaving others relatively underdeveloped; second, it needs to answer the questions of how technological advances occur and what the key processes and institutions involved are; and third, it has to present an analytical framework in which the role of technological change in regional economic growth is clearly explained.

## 5. *Redirecting Growth that Generates Garbage*

Tornell and Westermann (2000) find that financial liberalizations have been typically followed by lending booms that sometimes end in twin currency and banking crises. They highlight that lots of empirical work remains to be done in order to better characterize the mechanisms that underlie the boom-bust cycle, especially relating these cycles to the size of firms and across different sectors.

Paiche and Sterman (1992) also show that poor decision making in complex production systems can create pervasive booms and busts, where new products can have exponential sales increases, fuelling rapid growth, often leading to overcapacity, price wars, and bankruptcy.

Gai, Yao and Ye (2012) confirm the old adage, speed thrills but kills. They find that increasing the speed of trading from the microsecond level to the nanosecond level, lead to dramatic increases in message flow. However, the increases in message flow are due largely to increases in order cancellations without any real increases to actual trading volume. Spread does not decrease following increase in speed and market efficiency, in terms of price formation, does not improve. They find evidence that market depth decreases and short term volatility increases, probably as a consequence of more cancellations. Therefore, a fight for speed increases high-frequency order cancellation but not real high-frequency order execution. Increased cancellation generates more noise to the message flow. Low-frequency traders then subsidize the high-frequency traders because only executed trades are charged a fee.

Also, the exchanges continually make costly system enhancements to accommodate higher message flow, but these enhancements facilitate further order cancellations, not increases in trading volume. Investment in high frequency trading with sub-millisecond accuracy may provide a private benefit to traders without consummate social benefit; therefore, there may be an overinvestment in speed.

## IV. Conclusion

We looked at several illuminative analogies that have intuitively substantiated the coherence of the four pronged approach. We have also laid the groundwork for a formal approach towards proving these declarations. Given the breadth of the hypothesis, numerous revisions will be made and future versions of this study will continue to amplify these conclusions.

By considering each prong in isolation and then finally integrating the findings, we hope to establish different aspects of what would be crucial to increasing welfare and also what would be the limitations of any such recommendation. By recognizing the gap that exists between the fundamentals that drive the behavior of individuals or institutions, and the expected outcomes from their actions, primarily in the financial services, we hope to highlight how it becomes relatively straight forward to set incentives that can maximize welfare. If we start by reducing the size of institutions, it becomes easier to monitor them and a certain level of self-governance is also put into place. The other prongs then follow somewhat naturally and where there are deviations from what is desired, the reduced complexity, allows corrective mechanisms to be administered with less effort.

The dynamic nature of any social science system, like the financial services sector, means that the limited predictive ability of any awareness, will necessitate periodic reviews and programs then need to be prescribed in response to what is required. We certainly hope that this work will subsequently set the stage for an investigative methodology using the four pronged approach, specifically in economic activity and social sciences in general, and will lead to the formulation of appropriate policies. While this seems like a lofty ambition, the saying "aim for the stars or the skies and you will reach the tree tops" is the driving force behind this composition.

## V. References

Acharya, Viral V. "A theory of systemic risk and design of prudential bank regulation." *Journal of Financial Stability* 5.3 (2009): 224-255.

Acharya, V., Pedersen, L., Philippon, T., & Richardson, M. (2010). Measuring systemic risk.

Acharya, Viral, and Matthew Richardson, eds. *Restoring financial stability: how to repair a failed system*. Vol. 542. Wiley. Com, 2009.

Acs, Zoltan J., and Attila Varga. "Entrepreneurship, agglomeration and technological change." *Small Business Economics* 24.3 (2005): 323-334.

Acs, Zoltan J., and Attila Varga. "Geography, endogenous growth, and innovation." *International Regional Science Review* 25.1 (2002): 132-148.

Alexander, Carol. "Optimal hedging using cointegration." *Philosophical Transactions of the Royal Society of London. Series A: Mathematical, Physical and Engineering Sciences* 357.1758 (1999): 2039-2058.

Almgren, Robert, and Neil Chriss. "Optimal execution of portfolio transactions." *Journal of Risk* 3 (2001): 5-40.

Arestis, Philip, and Panicos Demetriades. "Financial development and economic growth: Assessing the evidence*." *The Economic Journal* 107.442 (1997): 783-799.

Arestis, Philip, Panicos O. Demetriades, and Kul B. Luintel. "Financial development and economic growth: the role of stock markets." *Journal of money credit and banking* 33.1 (2001): 16-41.

Avellaneda, Marco, and Sasha Stoikov. "High-frequency trading in a limit order book." *Quantitative Finance* 8.3 (2008): 217-224.

Barro, Robert J. "Economic growth in a cross section of countries." *The quarterly journal of economics* 106.2 (1991): 407-443.

Barro, Robert J. *Determinants of economic growth: a cross-country empirical study*. No. w5698. National Bureau of Economic Research, 1996.

Bastiat, Frederic. *The law*. Laissez Faire Books, 1968.

Basu, Parantap, and Alessandra Guariglia. "Foreign Direct Investment, inequality, and growth." *Journal of Macroeconomics* 29.4 (2007): 824-839.

Beck, Thorsten. "Bank competition and financial stability: friends or foes?." (2008).

Beck, Thorsten, and Asli Demirguc-Kunt. "Small and medium-size enterprises: Access to finance as a growth constraint." *Journal of Banking & Finance* 30.11 (2006): 2931-2943.

Beck, Thorsten, Asli Demirgüç-Kunt, and Ross Levine. "A new database on the structure and development of the financial sector." *The World Bank Economic Review* 14.3 (2000): 597-605

Beck, Thorsten, and Ross Levine. "Stock markets, banks, and growth: Panel evidence." *Journal of Banking & Finance* 28.3 (2004): 423-442.

Beck, Thorsten, Ross Levine, and Norman Loayza. "Finance and the Sources of Growth." *Journal of financial economics* 58.1 (2000): 261-300.

Beck, Thorsten, Asli Demirgüç-Kunt, and Ross Levine. "Financial institutions and markets across countries and over time-data and analysis." *World Bank Policy Research Working Paper Series, Vol* (2009).

Bodie, Kane, and A. Kane. "Marcus (2002). Investments."

Bonica, A., McCarty, N., Poole, K. T., & Rosenthal, H. (2013). Why Hasn't Democracy Slowed Rising Inequality?. *Journal of Economic Perspectives*, *27*(3), 103-24.

Borensztein, Eduardo, Jose De Gregorio, and Jong-Wha Lee. "How does foreign direct investment affect economic growth?." *Journal of international Economics* 45.1 (1998): 115-135.

Brealey, Richard A., Stewart C. Myers, and Franklin Allen. *Corporate finance*. Vol. 8. Boston et al.: McGraw-Hill/Irwin, 2006.

Caouette, John B., Edward I. Altman, and Paul Narayanan. *Managing credit risk: the next great financial challenge*. Vol. 2. John Wiley & Sons, 1998.

Chiu, J., Lukman, D., Modarresi, K., & Velayutham, A. (2011). High-frequency trading. *Standford, California, US: Stanford University*.

Chlistalla, M., Speyer, B., Kaiser, S., & Mayer, T. (2011). High-frequency trading. *Deutsche Bank Research*.

Christodoulakis, George A. "Bayesian Optimal Portfolio Selection: the Black-Litterman Approach." *Unpublished paper (available online at http://www. staff. city. ac. uk/~ gchrist/Teaching/QAP/optimalportfoliobl. Pdf)* (2002).

Ciccone, Antonio, and Robert E. Hall. *Productivity and the density of economic activity*. No. w4313. National Bureau of Economic Research, 1996.

Commission on the Measurement of Economic Performance and Social Progress, Stiglitz, J. E., Sen, A., & Fitoussi, J. P. (2009). Report by the commission on the measurement of economic performance and social progress.

Copeland, Laurence S. *Exchange rates and international finance*. Pearson Education, 2008.

Dempster, M. A. H., and C. M. Jones. "A real-time adaptive trading system using genetic programming." *Quantitative Finance* 1.4 (2001): 397-413.

De-Nicolo, Gianni. "Size, charter value and risk in banking: An international perspective." *EFA 2001 Barcelona Meetings*. 2001.

Derman, Emanuel, and Paul Wilmott. "The Financial Modelers' Manifesto." *SSRN: http://ssrn. com/abstract*. Vol. 1324878. 2009.

Diamond, Jared M., and Doug Ordunio. *Guns, germs, and steel*. New York: Norton, 1997.

Demirguc-Kunt, Asli, and Ross Levine. *Finance, financial sector policies, and long-run growth*. Washington, DC: World Bank, 2008.

De la Torre, Augusto, María Soledad Martínez Pería, and Sergio L. Schmukler. "Bank involvement with SMEs: beyond relationship lending." *Journal of Banking & Finance* 34.9 (2010): 2280-2293.

Easterly, William, and William Russell Easterly. *The elusive quest for growth: economists' adventures and misadventures in the tropics*. MIT press, 2001.

Elton, Edwin J., et al. *Modern portfolio theory and investment analysis*. John Wiley & Sons, 2009.

Engle, Robert F., and Clive WJ Granger. "Co-integration and error correction: representation, estimation, and testing." *Econometrica: journal of the Econometric Society* (1987): 251-276.

Fama, Eugene F., and James D. MacBeth. "Risk, return, and equilibrium: Empirical tests." *The Journal of Political Economy* (1973): 607-636.

Ferguson, Niall. *The ascent of money: A financial history of the world*. Penguin, 2008.

Gladwell, Malcolm. *The tipping point: How little things can make a big difference*. Hachette Digital, Inc., 2006.

Gladwell, Malcolm. *Outliers: The story of success*. Penguin UK, 2009.

Guilbaud, Fabien, and Huyen Pham. "Optimal high-frequency trading with limit and market orders." *Quantitative Finance* 13.1 (2013): 79-94.

Gujarati, Damodar N. "Basic econometrics, 3rd." (1995).

Hall, Robert E., and Charles I. Jones. "Why do some countries produce so much more output per worker than others?." *The quarterly journal of economics* 114.1 (1999): 83-116.

He, Guangliang, and Robert Litterman. "The intuition behind Black-Litterman model portfolios." *Available at SSRN 334304* (2002).

Hayek, Friedrich August. *The Road to Serfdom: Text and Documents--The Definitive Edition*. University of Chicago Press, 2009.

Hull, John. *Options, Futures, and Other Derivatives, 7/e (With CD)*. Pearson Education India, 2010.

Kashyap, R. (2014a). The Circle of Investment. *International Journal of Economics and Finance*, 6(5), 244-263.

Kashyap, R. (2014b). Dynamic Multi-Factor Bid–Offer Adjustment Model. *The Journal of Trading*, 9(3), 42-55.

King, Robert G., and Ross Levine. "Financial intermediation and economic development." *Capital markets and financial intermediation* (1993a): 156-189.

King, Robert G., and Ross Levine. "Finance and growth: Schumpeter might be right." *The quarterly journal of economics* 108.3 (1993b): 717-737.

King, Robert G., and Ross Levine. "Finance, entrepreneurship and growth." *Journal of Monetary economics* 32.3 (1993c): 513-542.

Levine, Ross. "Financial development and economic growth: views and agenda." *Journal of economic literature* 35.2 (1997): 688-726.

Levine, Ross. "Bank-based or market-based financial systems: which is better?." *Journal of financial intermediation* 11.4 (2002): 398-428.

Levine, Ross. "Finance and growth: theory and evidence." *Handbook of economic growth* 1 (2005): 865-934.

Levine, Ross, and David Renelt. "A sensitivity analysis of cross-country growth regressions." *The American economic review* (1992): 942-963.

Levine, Ross, and Sara Zervos. "Stock markets, banks, and economic growth." *American economic review* (1998): 537-558.

Levitt, Steven D., and Stephen J. Dubner. *Freakonomics Rev Ed LP: A Rogue Economist Explores the Hidden Side of Everything*. HarperCollins, 2006.

Mallaby, Sebastian. *More Money Than God: Hedge Funds and the Making of the New Elite*. Bloomsbury Publishing, 2011.

Marx, Karl, and Friedrich Engels. *The communist manifesto*. Yale University Press, 2012.

McCraw, Thomas K., ed. *Creating modern capitalism: How entrepreneurs, companies and countries triumphed in three industrial revolutions*. Harvard University Press, 1997.

Mises, Ludwig von. *The theory of money and credit*. London, 1980.

Murphy, Kevin M., Andrei Shleifer, and Robert W. Vishny. "The allocation of talent: implications for growth." *The quarterly journal of economics* 106.2 (1991): 503-530.

Natenberg, Sheldon. *Option volatility & pricing: advanced trading strategies and techniques*. McGraw Hill Professional, 1994.

Norstad, John. "The normal and lognormal distributions." (1999).

Olawale, Fatoki, and David Garwe. "Obstacles to the growth of new SMEs in South Africa: A principal component analysis approach." *African Journal of Business Management* 4.5 (2010): 729-738.

Olsson, Ola, and Douglas A. Hibbs Jr. "Biogeography and long-run economic development." *European Economic Review* 49.4 (2005): 909-938.

Paich, Mark, and John D. Sterman. "Boom, bust, and failures to learn in experimental markets." *Management Science* 39.12 (1993): 1439-1458.

Perold, Andre F. "The implementation shortfall: Paper versus reality." *Streetwise: the best of the Journal of portfolio management* (1998): 106.

Philippon, Thomas. *Financiers vs. Engineers: Should the Financial Sector be Taxed or Subsidized?*. No. w13560. National Bureau of Economic Research, 2007.

Philippon, Thomas, and Ariell Reshef. *Skill biased financial development: education, wages and occupations in the US financial sector*. No. w13437. National Bureau of Economic Research, 2007.

Philippon, Thomas, and Ariell Reshef. "Wages and Human Capital in the US Finance Industry: 1909–2006*." *The Quarterly Journal of Economics* 127.4 (2012): 1551-1609.

Popper, Karl R. "Poverty of Historicism." (2006).

Putterman, Louis, Valerie Bockstette, and Areendam Chanda. "States and Markets: the Advantage of an Early Start." (2001).

Rajan, Raghuram G., and Luigi Zingales. *Financial dependence and growth*. No. w5758. National Bureau of Economic Research, 1996.

Romer, Paul M. "Endogenous technological change." *Journal of political Economy* (1990): S71-S102.

Shlens, Jonathon. "A tutorial on principal component analysis." *Systems Neurobiology Laboratory, University of California at San Diego* (2005).

Smith, Adam. *An Inquiry into the Nature and Causes of the Wealth of Nations*. A. and C. Black, 1863.

Stiglitz, J., Sen, A., & Fitoussi, J. P. (2009). The measurement of economic performance and social progress revisited. *The Measurement of Economic Performance and Social Progress Revisited. Commission on the Measurement of Economic Performance and Social Progress, Paris*.

Sweeney, Joan, and Richard James Sweeney. "Monetary theory and the great Capitol Hill Baby Sitting Co-op crisis: comment." *Journal of Money, Credit and Banking* 9.1 (1977): 86-89.

Taleb, Nassim. *Fooled by randomness: The hidden role of chance in life and in the markets*. Random House Trade Paperbacks, 2005.

Taleb, Nassim Nicholas. *The Black Swan:: The Impact of the Highly Improbable Fragility*. Random House Digital, Inc., 2010.

Tornell, Aaron, and Frank Westermann. *Boom-bust cycles in middle income countries: Facts and explanation*. No. w9219. National Bureau of Economic Research, 2002.

Tuckman, B. "Fixed Income Securities---Tools for Today's Management." (1995).

Ye, Jiading Gai Chen Yao Mao, Chen Yao, and Jiading Gai. "The Externalities of High-Frequency Trading." *Available at SSRN 2066839* (2012).

Yusuf, Attahir. "Critical success factors for small business: perceptions of South Pacific entrepreneurs." *Journal of Small Business Management* 33 (1995): 2-68.

# Appendix

# VI. Research Questions

For the sake of completeness, we need to consider as many relevant questions as possible and undertake the effort to answer them. We need to mindful of the fact that it will be a mammoth effort to answer all the questions sketched below; hence we need to start with the questions that have an immediate connection with the financial services sector and welfare; and weave in the others as the various constraints dictate.

1. *Measure of Well Being*

    - Can we construct a smaller basket of goods and services indispensable for well-being? This basket would not be as comprehensive as the entire GDP of an economy, but would be representative of it and consider items like food, health care, education, real estate prices and disposable income.

- How can we combine this, basket of essential goods and services, with measures that include environmental; social; institutional and sustainability; indicators that give a more complete sense of well-being?

- How do we capture changes in this overall measure of well-being, across countries over time?

- As time passes, Luxuries become Necessities; hence we need to understand how our consumptions habits have changed over the years and across regions? We need to restrict our focus on the constituents of the welfare measure we have constructed.

2. *The Uncertainty Principle of the Social Sciences*

- How has the relationship between the measure of well-being that we have constructed and variables like inflation, interest rates, retail sales, unemployment rates, stock price indices, financial services changed over time and across different size populations?

- How have the above relationships changed as it applies to populations of different densities?

3. *The Responsibilities of Fiscal Janitors*

- What is the amount of money or money equivalents being handled by financial service institutions?

- The financial industry serves an important function by rewarding risk takers, supplying capital for new initiatives and providing liquidity. What is an accurate price for this service? Can this be answered by comparing the compensation of people in the financial services with compensations in other sectors? How crucial are capital markets activity and liquidity provided by the financial services sector to growth? Is there a level and composition most conducive for growth? How does the stage of economic development affect these associations? Different Asian economies are in varying stages of growth and will provide interesting case studies. We need to relate capital markets activity, the level of financial services development and compensations to GDP growth and our measure of well-being.

- Is there an optimal size of the financial sector versus the GDP? Countries have progressed from agricultural to industrial to financial economies. What is the long-term effect of this trend? How does this trend affect the measure we have created to capture well-being holistically?

- The role of interest rates in creating wealth and destroying it seems all too powerful. Though in reality the wealth effect due to interest rates is only in the record keeping or accounting and is rather superfluous. Is there a better mechanism for the determination of interest rates, given that our current knowledge in this area seems insufficient?

4. *The Need for Smaller Organizations*

- What is the amount of money or money equivalents being handled by financial service institutions of different sizes? How are the compensations across these size categories?

- How is the availability of different financing options for organizations of different sizes across time and across different regions?

- What is the relationship of our well-being measure to existing metrics that capture the strength of entrepreneurship in different regions and across time?

- What is the size of organizations and societies for optimal functioning? Was it absolutely crucial for big cities to develop for our technological capabilities to reach this advanced stage?

- How is the measure we have created for well-being across regions of different population densities?

- Do we need compact population centers despite the internet and easy access to knowledge in the public domain to stimulate innovation and technological progress?

- Can we have big cities with smaller size organizations? Or would smaller cities provide a better environment for smaller organizations?

5. *Redirecting Growth that Generates Garbage*

- How much waste, literally garbage, is generated by different countries with different levels of financial services growth, GDP, and with different levels of our well-being measure?

- How much waste is generated in terms of categories like plastic, paper or metal and also by different sectors of the economy? What is the connection of these segments to the financial services sector explicitly and implicitly?

- Increasing financial transactions might give the impression of growth but are they resulting in more welfare? We need to consider the relationship of financial transactions to the welfare measure we have created, across time and across different countries.

- The role of trading and the frequency of trading on the stability of the economic system and price volatility. The higher the frequency of trading, the more the liquidity that is provided to the market participants. The question remains, does the benefit in terms of greater liquidity provided by high frequency trading, offset the harm in other ways, by the increased volatility of prices and the need to devote a huge slice of society's resources in monitoring and handling this high frequency of money or money equivalents moving back and forth.

- What is the long-term necessity and influence of monetary and fiscal policy measures? GDP and Per-Capita income give a fair indication of living standards. This has led to increased focus on GDP growth in the short-term and excessive efforts to keep the growth rate high. The factors that affect long-term well-being are much more than the simple growth of goods and services. Hence, the increased short-term focus to stimulate the economy and keep high growth rates might not be required. It is true to a certain extent that increasing growth rates create more wealth and more wealth means better living conditions. However, in wealthy countries growth, though important, might not warrant excessive short-term attention. A measure of existing wealth distribution is more appropriate. In developing countries, instead of quarterly growth rates, something that measures whether the economy is on the right track might be helpful. We need to check the impact of fiscal and monetary policy measures used to stimulate growth on our

- overall measure of well-being and how this relationship varies across time and across different countries.
- Are the above problems simply manifestations of human nature, in seeking insurance or protection for the uncertainty in our lives and for the constant need we have to move from our current state to a, so called, better state. To find the answers to the above issues, do we need to probe further, into related areas of sociology?

## VII. Research Methodology

We need to probe the effects of the four prongs by exploring existing models and building new ones. The beauty and also the bane of measurement is that there is always something better or bigger and there is also something worse or smaller. This means we need to try to stick to relative measurements, where convenient, as opposed to absolute values. Once data from the below sources have been assembled into a suitable panel and various econometric checks have been done, it is a relatively mechanical process to keep spitting out different models, required to uncloak the patterns and relationships among the variables. The data gathering aspect, given the different types of data and length of history of the variables, becomes a highly important part of the entire study and given that we will not be able to collect all the required data at any given point in time; this will be an ongoing process. It is also highly improbable to think that we will be able to identify all the required data at any point; but the key takeaway is that we need to work with what we have and supplement the data set as new data elements become available or get constructed. A certain amount of creativity is warranted in selecting the right variables or modified forms of the variables in the analysis; but the method of operation will be quite unambiguous.

We need to collect information from the following sources

- Data released from the World Bank, http://data.worldbank.org
- Data from Global Financial, https://www.globalfinancialdata.com/index.html
- Data from Historical Statistics Organization, http://www.historicalstatistics.org
- Data from the Organization for Economic Co-operation and Development, (OECD), http://www.oecd.org
- Data from Bureau of Labor Statistics, http://www.bls.gov/data/
- Data on Garbage Generation, Food Production etc. from Nation Master, http://www.nationmaster.com/index.php
- Data from industry sources if possible; Thomson Reuters World Scope; Markit Purchasing Managers Index; Markit Intellectual Property data; Markit Credit Default Swap data; Bloomberg economic and security price data; and others.
- Data about patents from United States Patent and Trademark Office, http://www.uspto.gov
- Data from Global Entrepreneurship Monitor Cross Country Data to measure the level of entrepreneurship in each economy, http://www.gemconsortium.org/
- Data from New Business Registration offices, where available
- Data in the public filings of financial service firms

We need to collect the following variables and use it in our models and analysis.

1. *Measure of Well Being*

    - We need to construct an alternative to GDP that will use variables that show the increase in essentials like food, health care, education, real estate prices, and disposable income.

    - We need to supplement this proxy with variables for pollution; political stability; leisure time per day; vacation time; social support groups; sustainability index based on depletion of natural resources; unemployment security. This will be our measure of well-being.

    - Variables that capture the change in our consumption habits over the years, especially across the constituents of our welfare measure.

2. *The Uncertainty Principle of the Social Sciences*

    - We need not build any specific model to examine this prong but look at existing models relating growth and financial services.

    - We need to also look at existing models that relate different variables like inflation, interest rates, stock price indices, unemployment, or retail sales to growth.

    - We need to build models to relate the measure of welfare we have created to the variables in the prongs that follow and the other variables mentioned above.

    - The area of interest in this prong will be to assess the changes to the models over time and how the model parameters change as more individuals are exposed to it.

3. *The Responsibilities of Fiscal Janitors*

- Variables that measure the size of the financial services sector versus the rest of the economy.
- Variables that measure the compensation in financial services compared to other sectors over time.
- Variables that measure the change in compensation in the financial services sector when interest rates change and compare it to other sectors over time.
- Variables that will show the volume of money and money equivalents being moved around by financial firms.
- Variables that capture losses faced by financial service firms of different sizes.
- Variables that show the extent of risk taking in financial institutions of different sizes.
- Variables that capture stock market activity, credit markets activity and money supply changes.
- Variables that indicate the ratio of new funding provided to businesses versus secondary capital market transactions.
- We need to build a financial development index for different countries based on how long financial institutions have existed and the depth of financial services offered.
- We need to use case studies of the growth of financial services in different countries and see how that translates to welfare.

4. *The Need for Smaller Organizations*

- Variables that measure the size and number of individual financial service firms.

- Variables that measure the amount of money or money equivalents being handled by financial service institutions of different sizes. Here, we need to try and use variables that will show different types of financial instruments being handled.

- Variables that measure the compensation across financial service firms of different sizes and across different types of financial instruments.

- Variables that measure the number of small businesses that start in a given time period.

- Variables that measure the number of small businesses that close down in a given time period.

- Variables that measure the average length of time of existence of small businesses. This can refer to the gestation period of the business before it either flounders or moves into a different category.

- Variables that measure the extent of small businesses that get external funding.

- Variables that capture after what level of growth or after how many years of existence a small business will get access to external funding.

- Variables that capture the amount of capital being supplied to small businesses by financial firms of different sizes and categorized into the types of funding they are providing.

- Variables that show the number of new patents being developed.

- Variables that show the amount of new publications related to research in different areas.

- Variables that capture population growth and densities in different regions.
- Variables that relate the formation of small businesses to regions of different population growth and densities.
- Variables that relate patents and publications to regions of different population growth and densities.
- Variables that show the above metrics for firms of different sizes.
- Variables that will measure the change in these variables before and after the advent of social connectors like the internet.
- Variables that will capture whether the founder of a new business had worked elsewhere before staring the business. Where this is not available explicitly, we need to use age of the founder at the time of starting the business as a rough gauge.
- We need to build an index of entrepreneurship that shows how good an environment there is to start a new business and how that index has varied over time to show the culture of innovation prevalent in that region.

5. *Redirecting Growth that Generates Garbage*
    - Variables that capture the amount of garbage literally being generated by different countries with different growth rates and different levels of our welfare measure.
    - Variables that capture the types of waste being generated by different countries and by different sectors of the economy.
    - Variables that will indicate how much of the financial services sector is devoted to high frequency trading.

- Variables that relate financial service transactions to innovation within the industry and to our measure of well-being.
- Variables that capture the resources being devoted by exchanges and regulators to handle increased speed of transactions and to monitor the participants.
- Variables that measure the amount of news and content being generated related to GDP; to gauge monetary and fiscal policy impacts; to create financial services research; and resources devoted towards these efforts.
- Variables that capture new investments being made for different areas of the economy and different sub-segments within the financial services sector.
- Variables that capture how much insurance is being purchased and what types of insurance are being purchased.

After ensuring the data time series are clean and consistent, we need to perform stationarity checks and other econometric tests. We need to be careful to ensure that no biases among the data elements affect our results. We need to use panel data as opposed to cross-country regressions where enough data is available. The models then created will be tested for sensitivities across time. We need to create many smaller models and try to integrate the key variables from the smaller models into a larger framework.

As the first prong suggests, we need to build regression models that will show how the relationship between our welfare measure and the above variables changes across time and analyze the formation of pseudo-equilibrium states. We need to supplement these models with case studies across different countries and over time.